# Limited-angle TOF-PET for intraoperative surgical applications: Simulation Study

**S Sajedi[1], Y. Feng[1], H Sabet[1,*]**
[1]Department of Radiology, Massachusetts General Hospital, Harvard Medical School
* Corresponding author: HSabet@MGH.Harvard.edu

**Abstract:** In this work, we present modeling and imaging performance of a dual panel limited-angle TOF-PET system for intraoperative surgical applications using GATE monte carlo toolkit. Several detector parameters such as detector pixel dimensions, timing resolution and depth of interaction resolution along with tumor uptake ratio and phantom dimension are varied. Ultimately TOF-PET detector properties to achieve a specific imaging task are presented. To assess image resolution, we employed Simple Back Projection (SBP) reconstruction due to its fast speed compared to list-mode Maximum Likelihood Expectation Maximization (MLEM). We evaluated the quality of the reconstructed images using metrics contrast-to-noise ratio (CNR), contrast recovery coefficient (CRC), and signal-to-noise ratio (SNR). The purpose is to show effects of different detector parameters on the resolution of reconstructed images.

## 1 Introduction

The sentinel lymph node (SLN) concept, crucial for cancer staging, relies on gamma probes and near-infrared probes during surgery ([1],[2],[3],[2],[4],[5],[6]). However, it has been reported that identifying the SLN using current methods carries the risk of yielding false negative results ([7]), which fails to meet the recommendations of the American Association of Clinical Oncology (ASCO) ([8]), this highlights the need for more efficient tools of imaging. Despite the limitations of gamma probes, they are commonly used, prompting exploration of alternatives such as short-range positron particles from PET tracers like FDG. Time-of-flight (TOF) PET, with its enhanced image quality and reduced patient doses, emerges as a promising option. Intravenous (IV) administration of 18F-fludeoxyglucose (FDG) with positron emission tomography (PET) is a standard for tumor detection. Conventional whole-body PET (WB-PET) scanners, while effective for larger tumors, struggle with smaller ones due to low spatial resolution and sensitivity. To address this, intraoperative imaging devices have emerged, optimizing detection efficiency by placing detectors close to the tissue ([9], [10], [11], [12-15]).

In our previous study we demonstrated a limited angle tomography (LAT) TOF-PET proof of concept study using 3D printed hot lesion phantoms in the warm water background ([16, 17]). The study validated the experimental setup with the simulation model and the experiment results illustrated the effect of lesion to phantom activity (uptake) ratio and warm background water thickness over the image quality using the defined image quality parameters. In this work we extend the simulation study to show the effect of other scanner properties that could not be modified in the experimental study. In particular, we investigated the option of changing the crystal size, thickness, and Depth-of-Interaction (DOI) resolution as well as the TOF resolution of the scanner, and the phantom size and uptake ratio.

The choice of image reconstruction technique is important in limited angle system and should be adapted to the specific application. Incorporating TOF information in the reconstruction process can help reduce blurring caused by limited angular data ([18]). Various efforts have been undertaken to enhance the quality of images reconstructed from limited angular systems, including the use of penalized maximum-likelihood methods ([19]) and deep learning-based approaches ([20]). List-mode MLEM may outperform SBP in cases of limited angular data due to its ease of incorporating models or point spread function (PSF) kernels in the system matrix. However, this approach often comes with a higher computation cost which is not ideal for intraoperative setting. Given the imperative of on-the-fly image reconstruction with limited statistics in intra-operative imaging scenarios, the SBP method offers distinct advantages over list-mode MLEM. Our study focuses on investigating the effects of various system parameters on imaging resolution. We chose a simple back projection (SBP) method to generate the data, with the intention of exploring alternative image reconstruction methods to enhance image quality in future work. To elucidate the distinctions between various reconstruction techniques in LAT-PET, we conduct a comparison between an SBP-reconstructed image and one reconstructed using list-mode Maximum Likelihood Expectation Maximization (MLEM), with a coincidence time resolution (CTR) set to 100 ps. A comprehensive comparison is provided in the Methods section.

## 2 Materials and methods

### *2.1 Simulation setup*

In our recent work, we have shown through Monte Carlo simulation using GATE (GEANT4) that by bringing detector modules close to the patient, the detector solid angle and thus the geometrical sensitivity rapidly increases even with a small number of detector modules [21]. Our previous studies showed that by using ~1/3rd of the detector modules from a whole-body PET and rearranging them in a limited-angle body-contouring configuration, system sensitivity and resolution can be preserved. With these encouraging initial results, we further included flat panel detector geometries where the detector modules were placed in two parallel planes one right above the patient body and one below the surgery bed [16, 17]. The size of the probe is the main parameter to change the FOV and it should be carefully selected based on the application and the cost of the scanner. In this work we selected the probe and base detector sizes based on the designed phantom. Also, the distance between the probe and base detectors is selected as the minimum of reasonably achievable with patient and bed. In an ideal case the distance can be adjusted for each patient to have most solid angle coverage and image quality which at the same time will increase the recalibration burden and inaccuracy.

#### *2.1.1 Detector geometry*

The probe detector is supposed to be close to the region of interest and then it requires to have better spatial resolution, better DOI resolution, and higher count-rate capability. For the detector pitch we chose 2.2 mm pitch as the currently available very fine pitch silicone multiplier arrays (SiPM) as well as 3.3 mm pitch most commonly used arrays in the industry. For the base detector on the other hand, we chose two larger options as being farther from the ROI the pitch has less effect on the image resolution. In this regard we chose 4.2 mm pitch as an intermediate size and 6.2 mm largest readily available pitch in the SiPM market. With combination of probe crystal pitch sizes we implemented 4 detector geometries for simulation.

Since the Monte Carlo simulation is a time-consuming method, we tried to cover more parameters in a single simulation run in an efficient way. In this regard for each geometry, we chose 7 layers of 5 mm thick crystals connected seamlessly in the simulation geometry. By choosing arbitrary number of layers and discarding interactions over that layer in image reconstruction (which takes relatively less time) we could have different crystal thickness as 5, 10, 15, 20, 25, 30, and 35 mm. Obviously the thicker the crystal the higher the sensitivity of the scanner and the higher the cost. Although in the image reconstruction stage, the crystal thickness in probe and base can be chosen to be different, in this study we assumed that they are similar. Also, by merging the layer numbers in image reconstruction stage as (1) the DOI resolution of 5, 10, 15 mm or no DOI can be achieved depending on the crystal thickness.

$$layer\#_{ImRecon} = \left\lfloor \frac{layer\#_{Sim}}{n} \right\rfloor$$

Where the DOI resolution would be $n * 5\ mm$ and is used where the combination of accepted crystal layers and $n$ (DOI resolution) results a valid. Table 1 summarizes the valid combinations used in the image reconstruction. In practice it also makes sense for a scanner to have a better DOI resolution at probe than base to reduce the cost and still have similar performance. So, in addition to the 17 valid combinations that shows in table 1 where probe and base have similar configuration there are 8 more possibility that base has no DOI and probe has more than one DOI level.

Table 1. DOI resolution in the image reconstruction based on the chosen layers and merging combination

| DOI Res. | Crystal thickness (mm) | | | | | | |
|---|---|---|---|---|---|---|---|
| | 5 | 10 | 15 | 20 | 25 | 30 | 35 |
| 5 mm (n=1) | valid | valid | valid | valid | valid | valid | valid |
| 10 mm (n=2) | - | valid | - | valid | - | valid | - |
| 15 mm (n=3) | - | - | valid | - | - | valid | - |
| 20 mm (n=4) | - | - | - | valid | - | - | - |
| 25 mm (n=5) | - | - | - | - | valid | - | valid |
| 30 mm (n=6) | - | - | - | - | - | valid | |
| 35 mm (n=7) | - | - | - | - | - | - | valid |

Although the GATE code provides digitizer that can simulate different timing performance of detectors, we chose to not to use it and implement the time blurring of the events in the image reconstruction code, based on the target coincidence time resolution of the scanner. In order to this a Gaussian blurring is applied to each single event prior the reconstruction with standard deviation as:

$$\sigma = \frac{CTR_{(FWHM)}}{2.355 * \sqrt{2}}$$

Where $CTR_{(FWHM)}$ is the target scanner's CTR. In this study 5 different target CTRs for the scanners is chosen as 10 ps, 50 ps, 100 ps, 200 ps, 300 ps. These numbers span from the low cost to the current state-of-the-art technology and potential future technologies.

### *2.1.2 Phantom*

The designed phantom consists of 4 regions of hot spheres and a central cold sphere all placed inside a tank 2 cm away from the top. The hot spheres have diameters of 2, 4, 6, and 8 mm with a gap of the same size between them. Figure 1 left shows the hot and cold spheres from coronal view. The warm background phantom consists of water with 5300 Bq/cc activity chosen to be similar to the clinical PET diagnostic imaging protocols.

While the location of the hot and cold spheres in respect to the probe and base, is fixed and their center is 2 cm below the water surface in the phantom and 2.5 mm below the probe detector entrance (see Fig 1 right), the thickness of the water phantom can be changed to explore the effect on the image quality. We chose 25 cm as the body thickness for a normal patient in chest area as well as 12 cm as an intermediate size for normal breast, and 5 cm for a compressed breast tissue.

The uptake of the radiotracer in the tumor is a biological property of the tracer and irrelevant of the scanner properties but can make a significant difference in the image quality. To better understand the scanner performance under different clinical scenarios we chose to have 4 different uptake ratios between the hot spheres as the lesions and the warm water background. The selected ratios are 2.5, 5, 10, and 20. Each ratio is combined with three phantom thicknesses creating 12 different phantoms for one geometry as the reference (2.2 mm SiPM at probe and 4.2 mm SiPM pitch at base, see section 2.1.1). for other 3 detector geometries the uptake ratio of 10 was chosen to be constant for different combination of phantom thicknesses creating 9 more simulation scenarios.

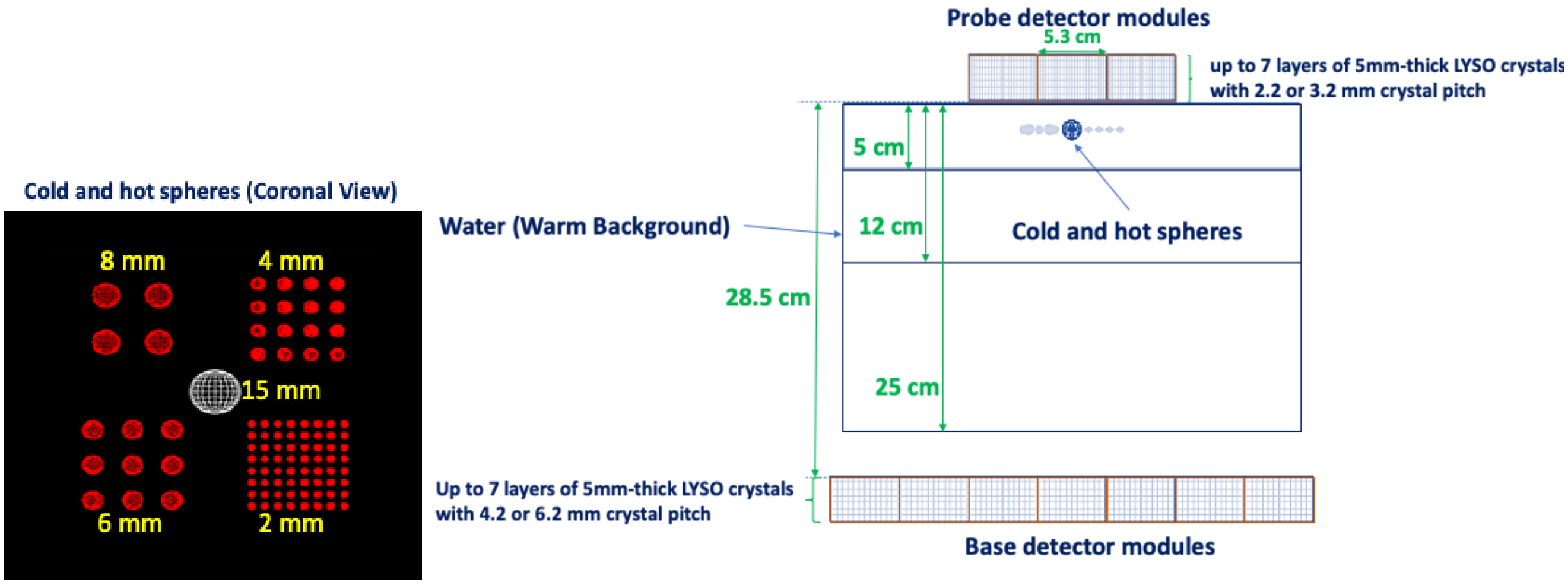

Fig. 1. Left) Coronal view of the created lesion phantom in GATE with hot spheres depicted in red color and cold sphere in white color. Right) The lateral view of the geometry with overlayed three different depths of the background water phantom. Three phantom size and two different crystal pitches for both probe and base detectors are simulated (12 runs in total) and, on each run, 7 layers of 5 mm thick LYSO crystals are simulated in order to select desired crystal thickness and DOI resolution in the image reconstruction code. The timing resolution is also simulated in the image reconstruction code.

Each three phantom thicknesses are combined with 4 detector geometries described in 2.1.1 resulting 12 different simulation scenarios in GATE.

### *2.2 Image quality metrics*

After reconstruction, we evaluated the image quality using the contrast recovery coefficient (CRC), signal-to-noise ratio (SNR), and contrast-to-noise ratio (CNR). CRC is described in the NEMA standard to measure the quality of hot rods in the scan of micro-Derenzo phantoms. It represents how much of the real activity is recovered in the reconstructed image. We modified the CRC to reflect the use of hot spheres instead of rods (with no slice summation) as follow:

$$CRC = \frac{\frac{Mean_{HotSphere}}{Mean_{Background}} - 1}{\frac{Activity_{HotSphere}}{Activity_{Background}} - 1}$$

where "*Activity*" refers to the activity of the phantom and background in the same region of interest (ROI) where the "Mean" is calculated. CNR shows the detectability of the hot spheres and is defined as the ratio of the image contrast to the noise in the background:

$$CNR = \frac{Mean_{HotSphere} - Mean_{Background}}{Std_{Background}}$$

and SNR is calculated using:

$$SNR = \frac{\frac{Mean_{HotSphere}}{Mean_{Background}} - 1}{\sqrt{\left(\frac{Std_{HotSphere}}{Mean_{HotSphere}}\right)^2 + \left(\frac{Std_{Background}}{Mean_{Background}}\right)^2}}$$

where "*Std*" refers to the standard deviation of pixel values. The ROI for the hot spheres is obtained based on the accurate alignment of the phantom and detectors and not from the pixel values. For the background, we used two circular shapes with a 15 mm diameter in the same image plane with center coordinates at 40 mm and -35 mm from the center of the FOV at the horizontal axis and center at the vertical axis. For the hot spheres, the parameters for each of the two spheres with the same size are calculated separately, and the mean and standard error of two values are calculated. Since the two smaller sphere sizes are not visualized and the parameters do not represent a meaningful value to compare, here we report sphere sizes of only 5.7 mm, 7.8 mm, and 9.4 mm.

### *2.3 Image reconstruction*

It is well established that the conventional filtered back-projection (FBP) method will induce artifact in the reconstructed image when we don't have full 2pi view-angle covered by LORs. Furthermore, the rebinning methods used to map the oblique planes to direct planes, would be a compromise of error and sensitivity based on the maximum acceptable ring difference. The 3D image reconstruction methods are developed to obtain the best sensitivity and spatial resolution trade-off. The 3D filtered back-projection method is developed to deal with this problem and tries to compensate for the limited view-angle coverage in phi direction (due to limited axial coverage). Many efforts have been carried out to optimize the results ([22, 23]). In this work, where there is a same angular coverage in theta and phi direction, the same methodology may be used to reconstruct image using the 3D filtered back-projection.

The iterative image reconstruction methods on the other hand, can model arbitrary geometry and more intrinsic models in the detector within same algorithm structure. Because of this versatility, it is usually the first choice where the geometry of imaging is not standard full ring.

Intra-operative imaging requires on-the-fly image reconstruction with minimum latency to be applicable during a normal operation procedure. This adds a strict constraint to the image reconstruction method to be used. For instance, the iterative image reconstruction requires to iterate over the acquired data at least a few times to converge to a reasonably good image quality, which means, it should start after the data acquisition is done at least for a predefined shorter imaging time interval. It may make the usage of iterative image reconstruction method more challenging to be useful in a real intra-operative application. Furthermore, the less the computational budget required for image reconstruction the less latency imposed with the same computational power. The exact discussion about the computational power requires detail knowledge of the method and implementation, but with the advances in the GPU based reconstruction systems this may be less challenging where the cost of increasing the computational budget is significantly less than imaging detectors and readout system.

The criteria of choosing the reconstruction method in this work is slightly different where it is not going to be used in real application where it need on-the-fly image reconstruction, but similarly it should be fast enough so we can perform image reconstruction for large number of scanner designs with different parameters. More importantly the analytical image reconstruction methods usually show less affected by reconstruction parameters and make a better apple to apple comparison because of their linearity.

In this regard we chose the simple-back projection with TOF information as the simplest method which is also used in our previous work for experimental proof of concept. To know how this method can affect the comparisons in this work, we compared this method with maximum likelihood expectation maximization (MLEM) method which is the mostly used iterative method for PET image reconstruction. The list mode MLEM was chosen because of relatively low counts in this application and also it does not need TOF binning. One iteration of the list mode MLEM reconstruction can be expressed as:

$$\lambda_j{}^{l+1} = \frac{\lambda_j{}^{l}}{S_j} \sum_i T_{ij} * \frac{1}{\sum_{j'} T_{ij'} \lambda_{j'}{}^{l}} ,$$

$\lambda_j{}^{l}$ is the estimated image voxel *j* at *l*th iteration, S is the sensitivity matrix and is normalized to 1, T is the PET system matrix and is calculated on the fly. One element in T corresponds to the possibility of one event emitted from voxel *j* being detected by the line of response (LOR) defined by one coincidence *i*. The timing information is added to T and is calculated as a single Gaussian kernel. The correction of random and scattering is not considered in this comparison. All coincidences detected were used in the MLEM reconstruction.

Figure 2 shows the comparison of the SPB used in this work with the implemented MLEM method with $4.8\times10^{7}$ counts in total. The reconstruction volume is divided into 256×256×1 voxels with one voxel measuring 0.5 $mm^3$. The lesion to background ratio is 20, warm background depth is 5 cm. For image reconstruction 100ps CTR is modeled and 7 layers with 5 mm-thick crystal are used in the reconstruction. The current MLEM reconstruction is paralleled in the data space and takes 1420 s per iteration on 2.7 GHz CPU with 128 threads. The reconstructed image using MLEM at 50th iteration is shown in figure 2 (left).

As expected with simple back projection the reconstructed image has lower noise as the ramp filter (or other ramp like filters) is not applied to each projection. The lower noise level will make CNR and SNR values much better based on their definition and the CRC difference is not significant enough to be comparable. The visual image quality also shows less noise with better detectability of 2 mm spheres, which is the goal of the scanner design.

While the general comparison of two image reconstruction methods requires comprehensive investigation for final application, for this work as the purpose is to compare different geometries and scanner properties, and we didn't see significant degradation of image quality, we chose to use the SBP method were the linearity of the method and being less dependent to detector modeling and iteration number for different geometries, is the key for systematic investigation.

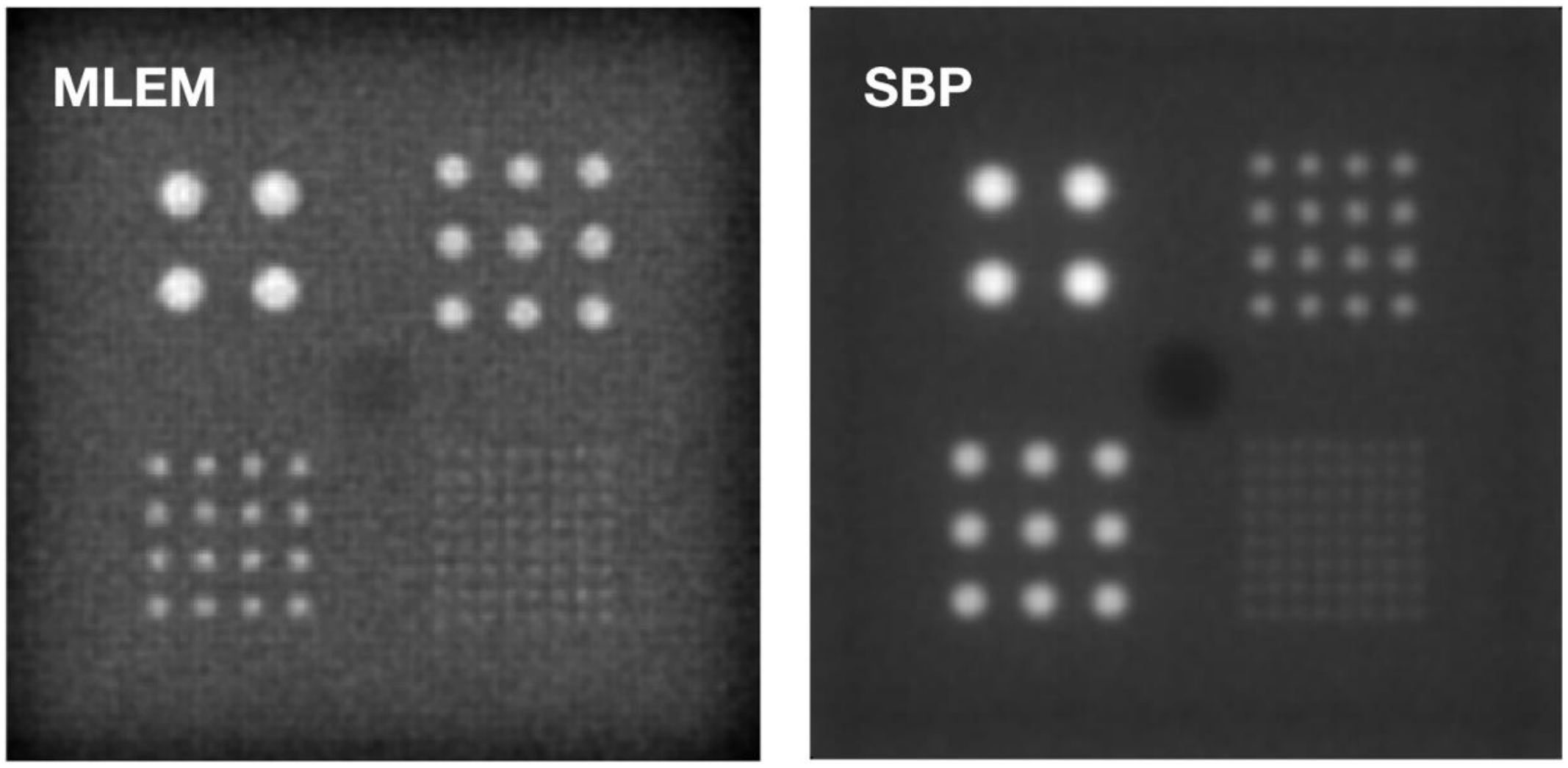

Fig. 2. Comparison of the reconstruction methods with (left) MLEM after 50 iterations and (right) SBP. Both methods modeled 100 ps CTR and used 0.5 mm image voxel size, 7 layers of 5 mm thick detector crystals in both probe and base, and 2.2 mm crystal pitch at probe and 4.4 mm pitch at base. The warm background water thickness is 5 cm and lesion to background ratio is 20.

## 3 Results

For each of 21 simulation scenarios the coincidence data is collected for 60 second data acquisition. 5 CTR values and 25 crystal thickness-DOI combinations (see Table 1) are used to create 125 reconstructed images for each simulation scenario. The results shown in Figs 2-10 here is a summary of 2625 reconstructed images changing only one parameter at the time and measuring the reconstructed image parameters.

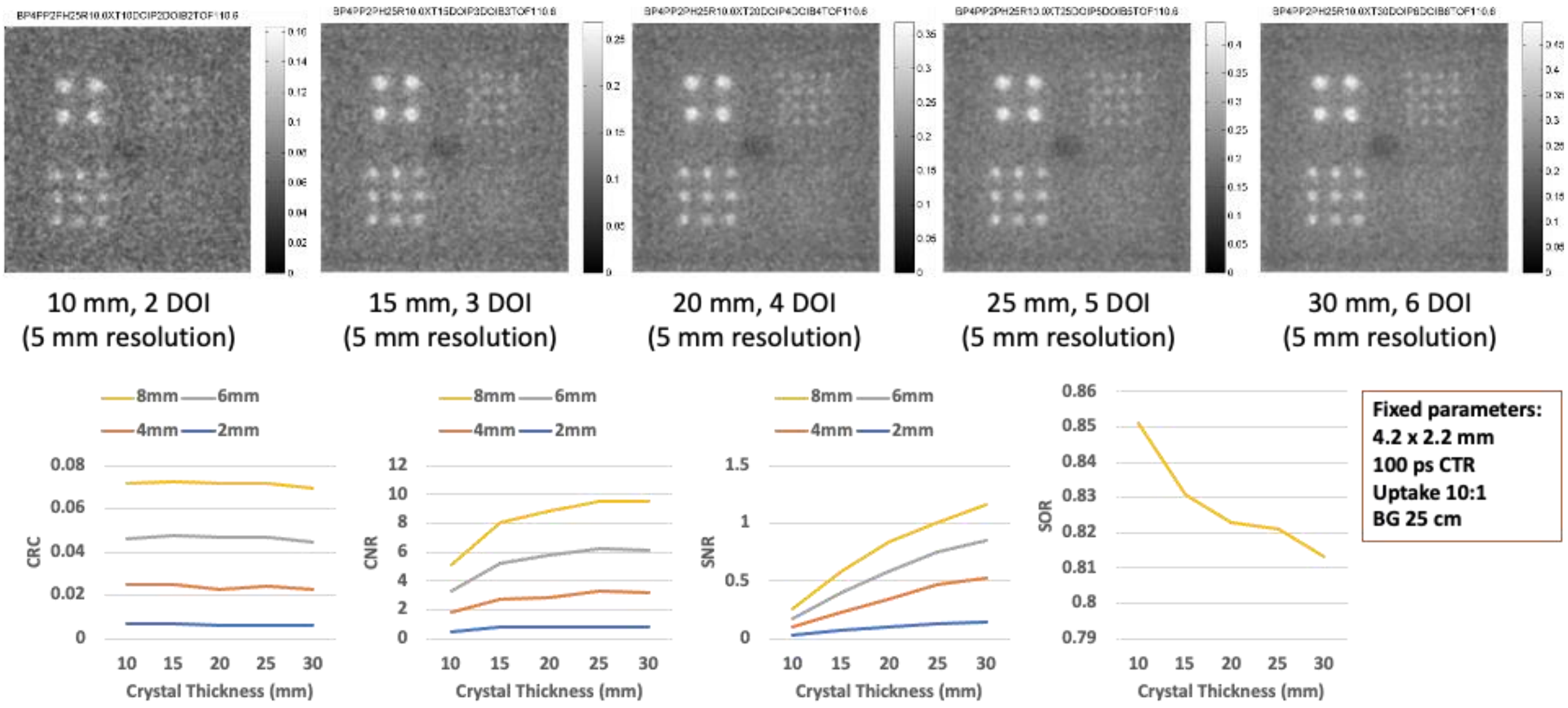

Fig. 3. Effect of crystal thickness.

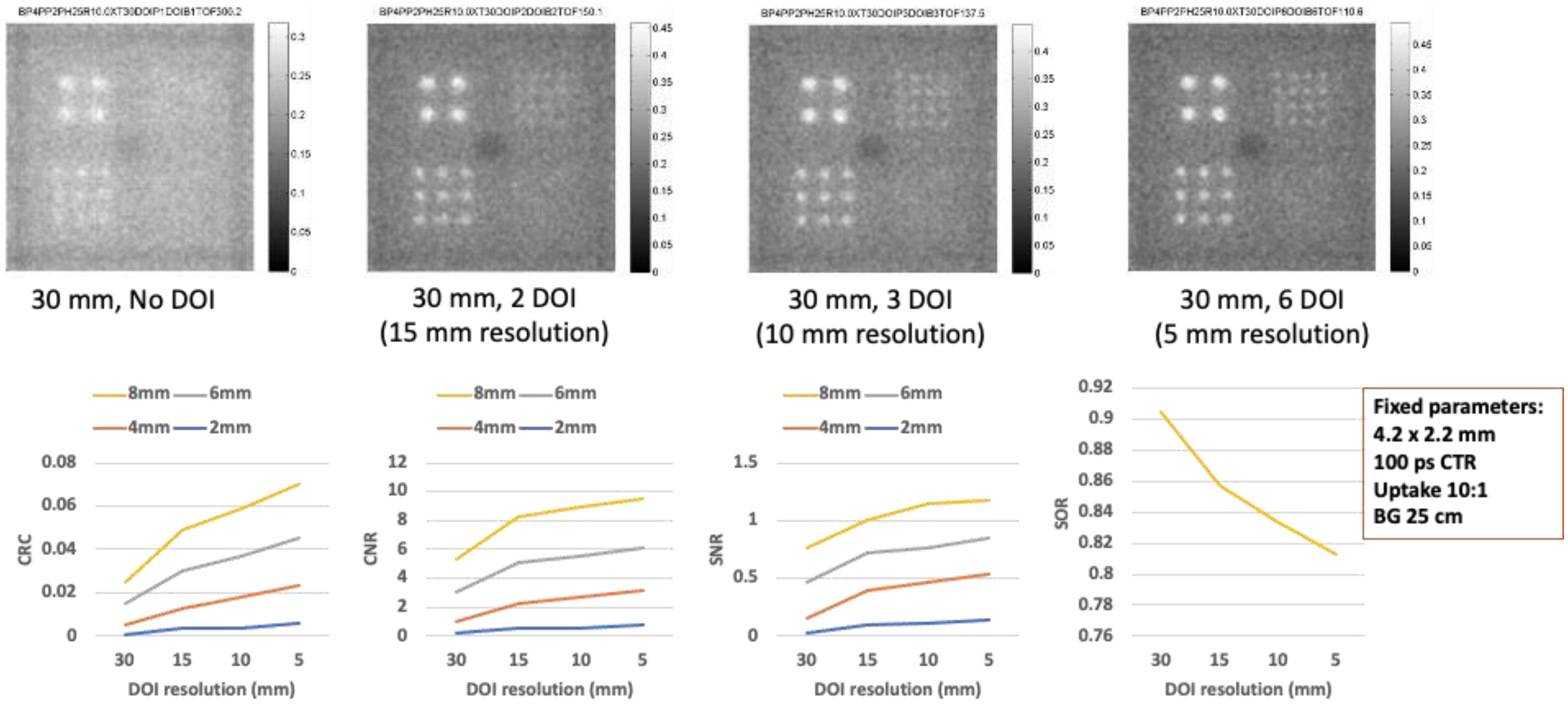

Fig. 4. Effect of DOI.

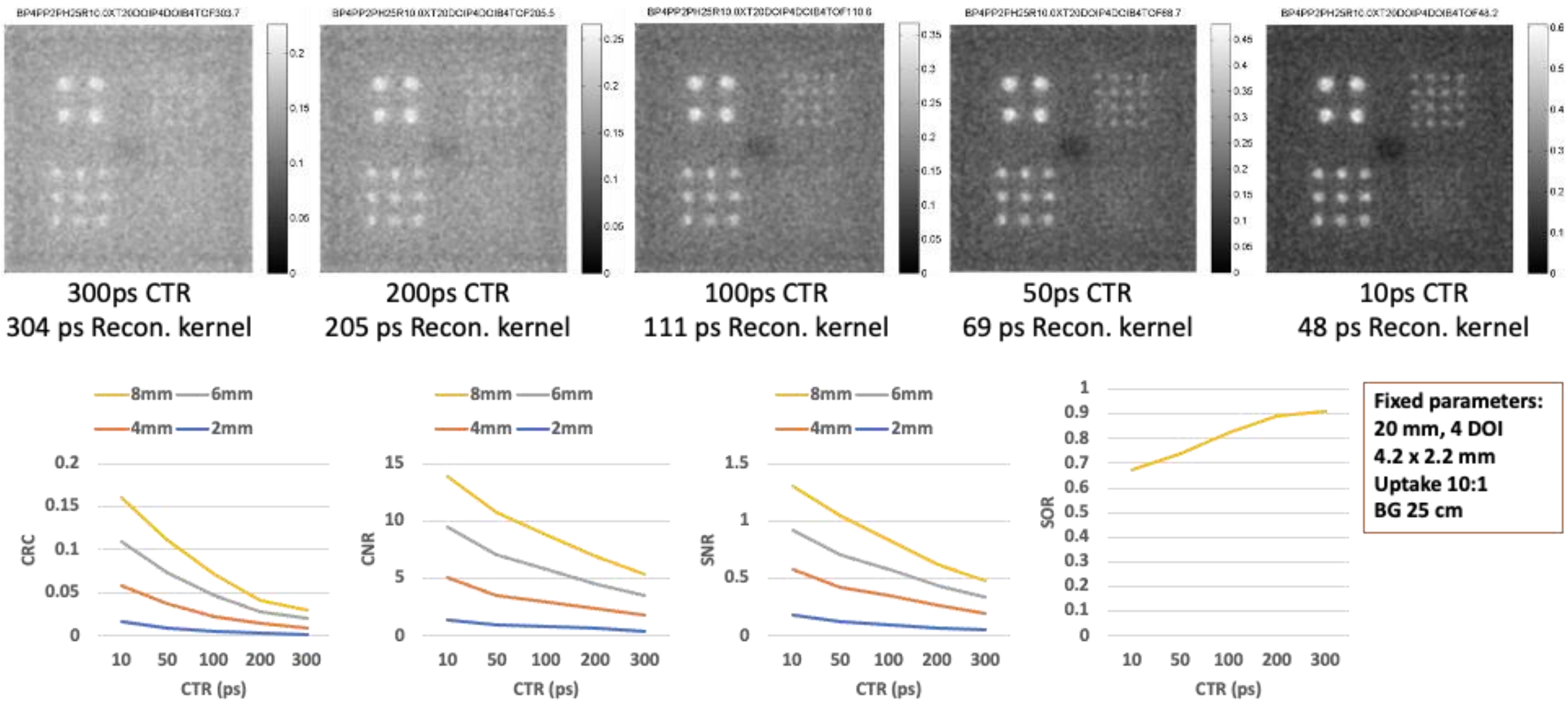

Fig. 5. Effect of TOF.

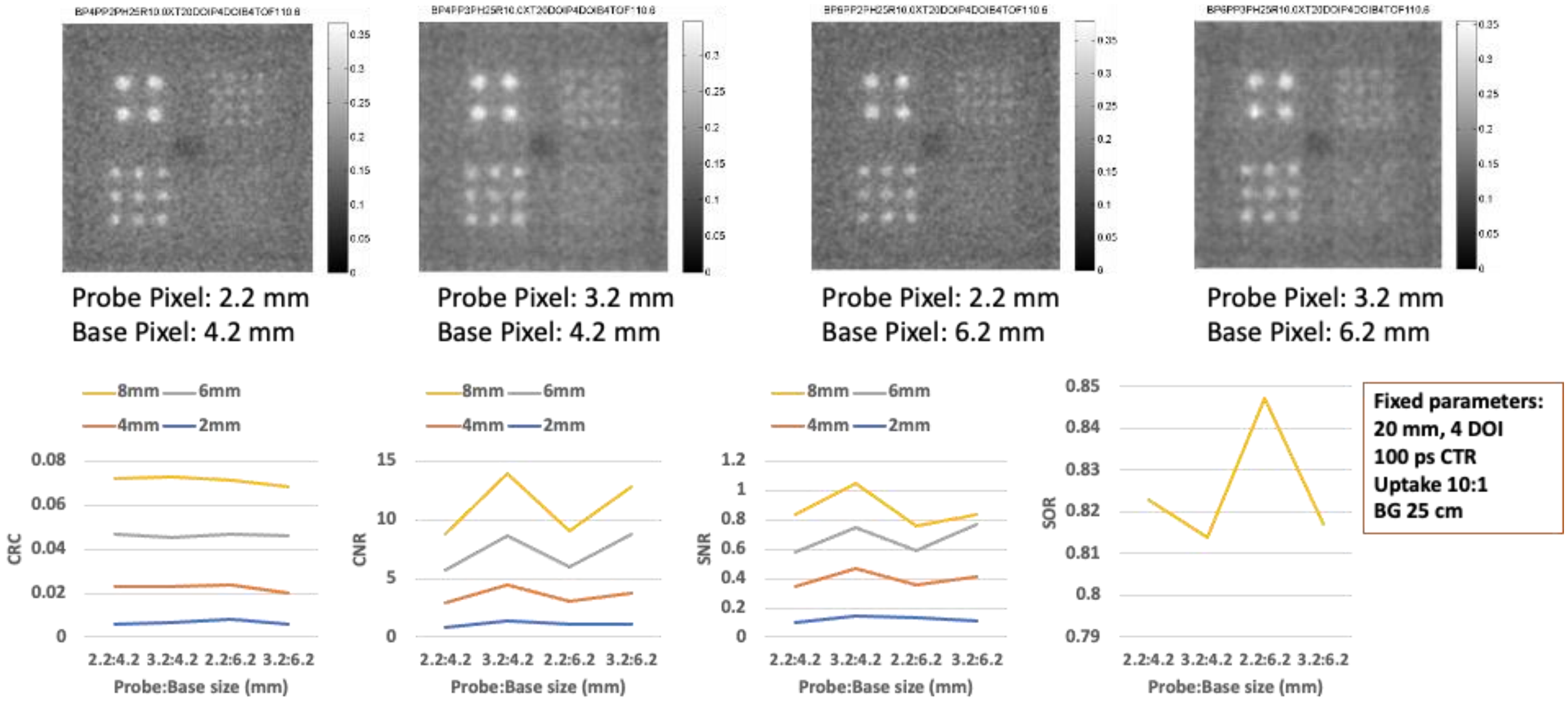

Fig. 6. Effect of Pixel Pitch.

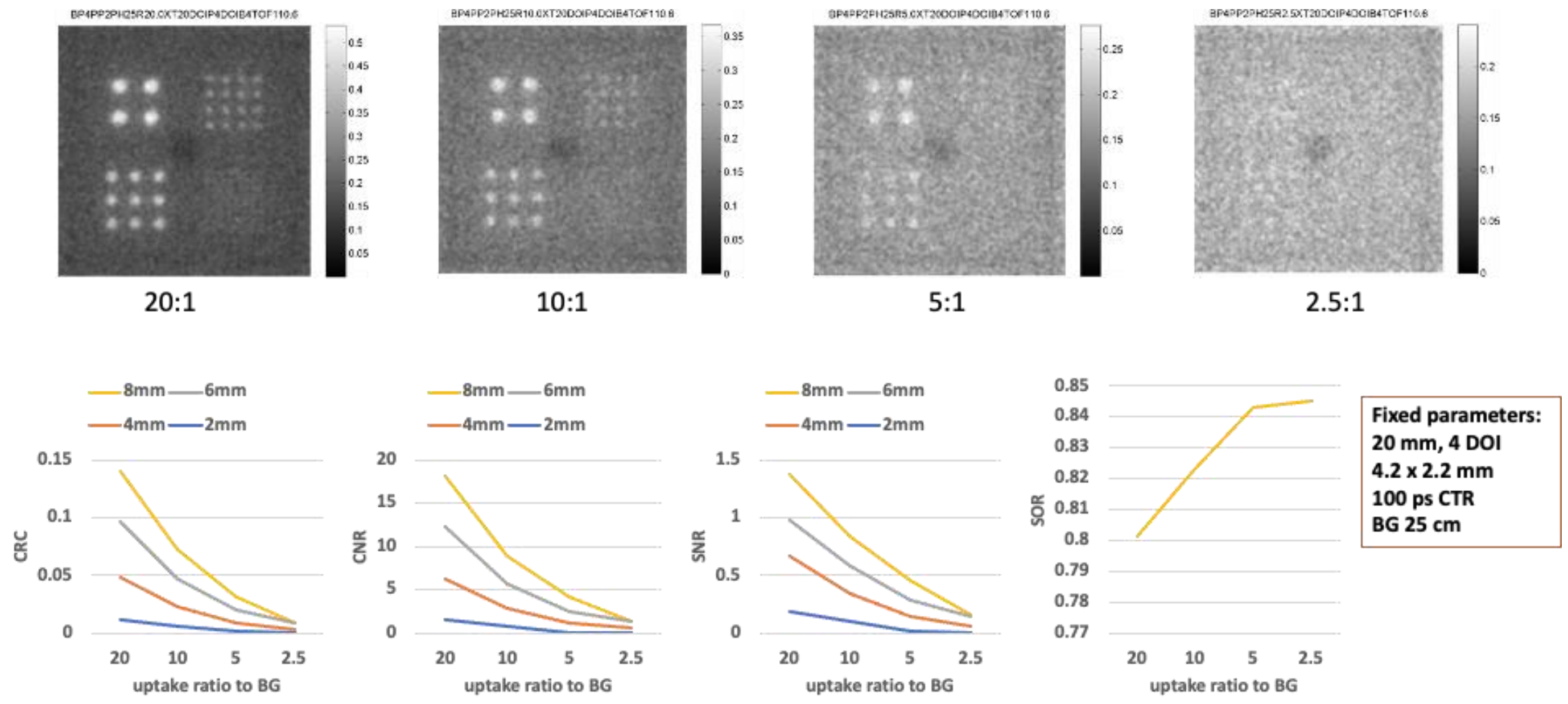

Fig. 7. Effect of uptake ratio

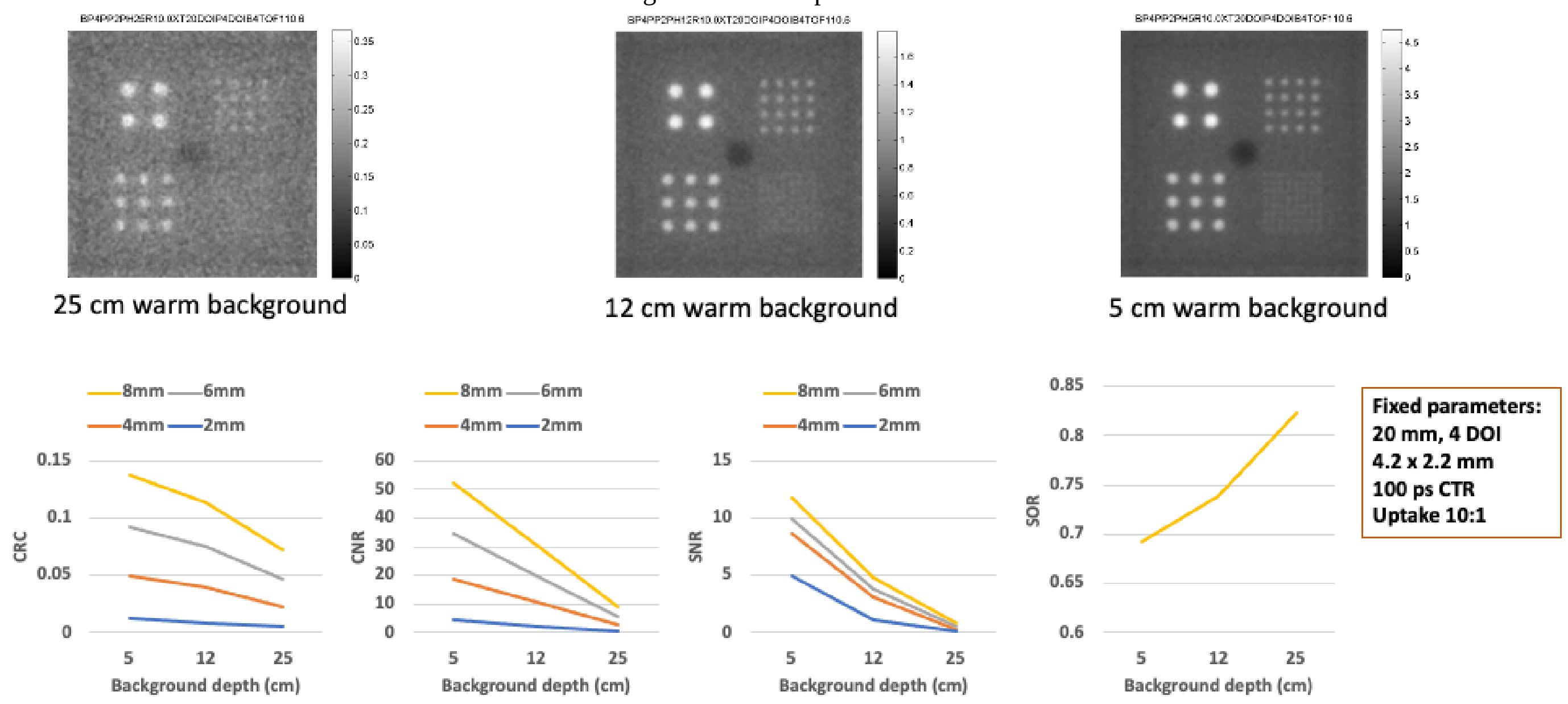

Fig. 8. Effect of Background thickness

Required scanner CTR with 1 minute scan, 20 mm crystal and 5 mm DOI resolution

| 2 mm Spheres | Ratio 2.5:1 | Ratio 5:1 | Ratio 10:1 | Ratio 20:1 |
|---|---|---|---|---|
| 25 cm | Not Detectable | Not Detectable | Not Detectable | 10 ps Border line |
| 12 cm | Not Detectable | Not Detectable | 50 ps CTR | 200 ps CTR |
| 5 cm | Not Detectable | Not Detectable | 200 ps CTR | 300 ps CTR |

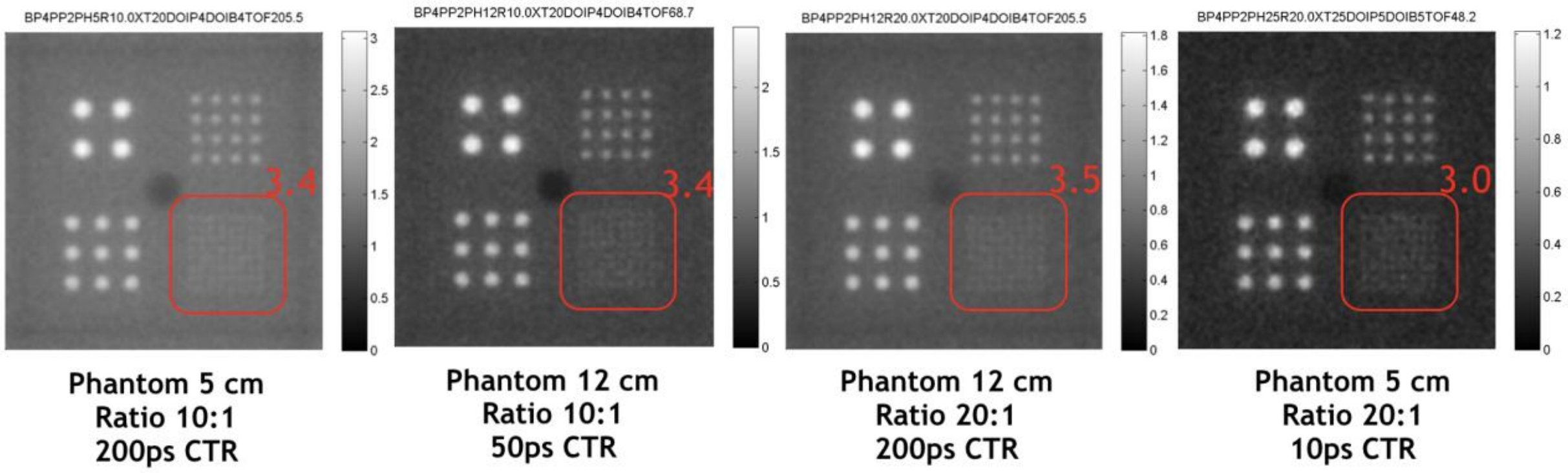

Fig.9. Required scanner CTR with 1-minute scan, 20 mm crystal and 5 mm DOI resolution for 2 mm rods.

Required scanner CTR with 1 minute scan, 20 mm crystal and 5 mm DOI resolution

| 4 mm Spheres | Ratio 2.5:1 | Ratio 5:1 | Ratio 10:1 | Ratio 20:1 |
|---|---|---|---|---|
| 25 cm | Not Detectable | Not Detectable | 50 ps CTR | 300 ps CTR |
| 12 cm | Not Detectable | 100 ps CTR | 300 ps CTR | 300 ps CTR |
| 5 cm | 50 ps CTR | 300 ps CTR | 300 ps CTR | 300 ps CTR |

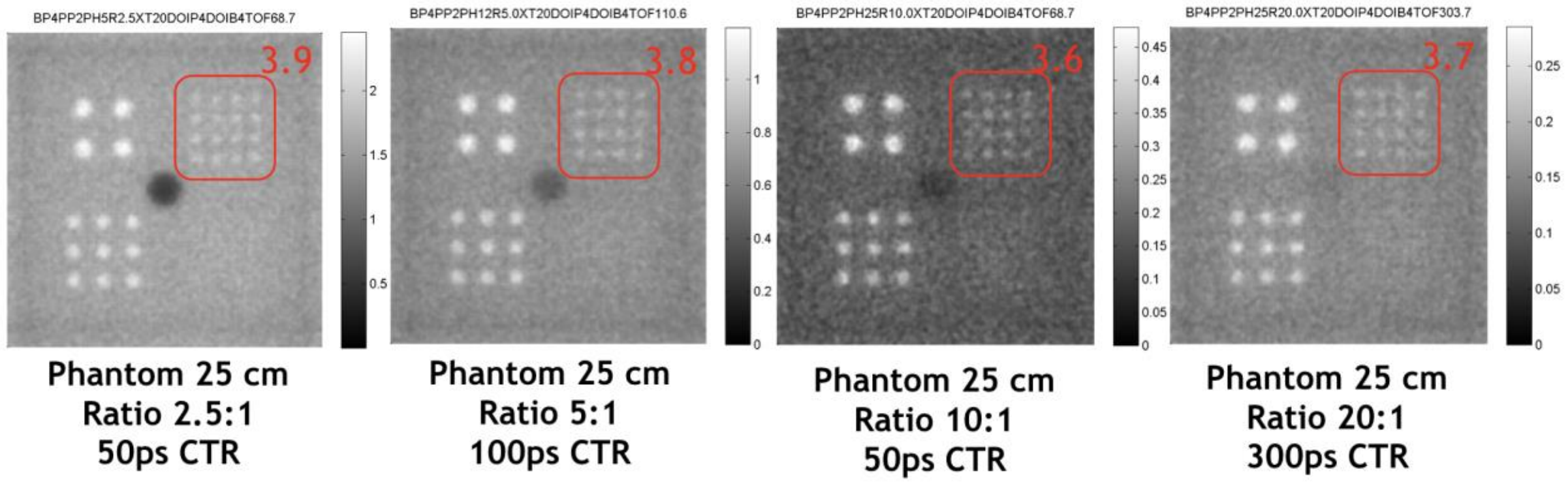

Fig. 10. Required scanner CTR with 1-minute scan, 20 mm crystal and 5 mm DOI resolution for 4 mm rods.

## 4 Discussion and Conclusion

### *4.1 Detectability*

The detectability of system is defined by the CNR and the threshold of the CNR is obtained from ROC curves with highest AUC for reconstructed image in the specific view and cross-section where all hot lesions are best visible. The tables in Figures 9 and 10 provide a summary of the detectability of the 2 mm and 4 mm rods with various background ratios and background thickness. Results demonstrate that at a background thickness of 25 cm, 2 mm rods with a 20:1 ratio can only be resolved with a 10 ps CTR, while 4 mm rods with a 10:1 ratio are detectable with a 50 ps CTR, and 4 mm rods with a 20:1 ratio are detectable with a 300 ps CTR. At a background thickness of 12 cm, 2 mm rods with a 10:1 ratio are detectable with a 50 ps CTR, and 20:1 ratio rods are resolvable with a 200 ps CTR. Furthermore, 4 mm rods with a 5:1 ratio can be resolved with a 100 ps CTR. At a background thickness of 5 cm, 2 mm rods with a 10:1 ratio can be resolved with a 200 ps CTR, while all 4 mm rods with ratios greater than 5:1 can be resolved with a 300 ps CTR. However, 4 mm rods with a 2.5:1 ratio can only be resolved with a 50 ps CTR.

### *4.2 Conclusion*

The results indicate that using a 100 ps CTR, it is possible to detect 4 mm rods with a 12 cm thickness of warm background, even with a low uptake ratio (5:1). Additionally, in the same thickness, 2 mm rods with a larger uptake ratio (10:1) can be resolved with a 50 ps CTR.

All the images shown in Figures 3-10 were obtained using SBP, list mode MLEM reconstruction was not applied due to computational constraints. As shown in Figure 2, the images reconstructed with SBP exhibit a smoother quality compared to MLEM reconstruction, primarily caused by the limited statistics. No resolution recovery methods were applied in this study; however, we intend to incorporate Point Spread Function (PSF) modeling into the MLEM reconstruction to enhance imaging resolution in future iterations.

In future work, we explore that new technologies such as the large area picosecond photodetector [24], fast electronics [25], novel layered detector [26], and GPU accelerated image reconstruction [27] and machine learning [28, 29]not only to boost image quality but also to facilitate sub-minute image generation for intraoperative surgical applications.

## References

1. Borgstein, P.J., et al., *Sentinel lymph node biopsy in breast cancer: guidelines and pitfalls of lymphoscintigraphy and gamma probe detection.* Journal of the American College of Surgeons, 1998. **186**(3): p. 275-283.
2. Krag, D.N., et al., *Surgical resection and radiolocalization of the sentinel lymph node in breast cancer using a gamma probe.* Surgical oncology, 1993. **2**(6): p. 335-340.
3. Ae, G., *Lymphatic mapping and sentinel lymphadenectomy for breast cancer.* Ann surg, 1994. **220**: p. 391-398.
4. Morton, D.L., et al., *Technical details of intraoperative lymphatic mapping for early stage melanoma.* Archives of surgery, 1992. **127**(4): p. 392-399.
5. Kaviani, S., et al., *Design and development of a dedicated portable gamma camera system for intra-operative imaging.* Physica Medica, 2016. **32**(7): p. 889-897.
6. Kaviani, S., et al., *Development and characterization of a compact hand-held gamma probe system, SURGEOGUIDE, based on NEMA NU3-2004 standards.* Journal of Instrumentation, 2016. **11**(12): p. T12004.
7. Pesek, S., et al., *The false-negative rate of sentinel node biopsy in patients with breast cancer: a meta-analysis.* World journal of surgery, 2012. **36**: p. 2239-2251.
8. Lyman, G.H., et al., *American Society of Clinical Oncology guideline recommendations for sentinel lymph node biopsy in early-stage breast cancer.* Journal of clinical oncology, 2005. **23**(30): p. 7703-7720.
9. Sajedi, S., H. Sabet, and H.S. Choi, *Intraoperative biophotonic imaging systems for image-guided interventions.* Nanophotonics, 2018. **8**(1): p. 99-116.
10. Sabet, H., B.C. Stack, and V.V. Nagarkar, *A hand-held, intra-operative positron imaging probe for surgical applications.* IEEE Transactions on Nuclear Science, 2015. **62**(5): p. 1927-1934.
11. Daghighian, F., et al., *Intraoperative beta probe: a device for detecting tissue labeled with positron or electron emitting isotopes during surgery.* Medical physics, 1994. **21**(1): p. 153-157.
12. Sabet, H., et al., *A method for fabricating high spatial resolution scintillator arrays.* IEEE Transactions on Nuclear Science, 2013. **60**(2): p. 1000-1005.
13. Sabet, H., B.C. Stack, and V.V. Nagarkar. *A novel intra-operative positron imager for rapid localization of tumor margins*. in *Medical Imaging 2014: Physics of Medical Imaging*. 2014. SPIE.

14. Spadola, S., et al., *Design optimization and performances of an intraoperative positron imaging probe for radioguided cancer surgery.* Journal of Instrumentation, 2016. **11**(12): p. P12019.
15. Stendahl, J.C., et al., *Prototype device for endoventricular beta-emitting radiotracer detection and molecularly-guided intervention.* Journal of Nuclear Cardiology, 2022. **29**(2): p. 663-676.
16. Sajedi, S., et al. *Limited-angle TOF-PET for intraoperative surgical applications: latest results*. in *2020 IEEE Nuclear Science Symposium and Medical Imaging Conference (NSS/MIC)*. 2020. IEEE.
17. Sajedi, S., et al., *Limited-angle TOF-PET for intraoperative surgical applications: proof of concept and first experimental data.* Journal of Instrumentation, 2022. **17**(01): p. T01002.
18. Gravel, P., Y. Li, and S. Matej, *Effects of TOF resolution models on edge artifacts in PET reconstruction from limited-angle data.* IEEE transactions on radiation and plasma medical sciences, 2020. **4**(5): p. 603-612.
19. Zhang, H., et al., *Penalized maximum-likelihood reconstruction for improving limited-angle artifacts in a dedicated head and neck PET system.* Physics in Medicine & Biology, 2020. **65**(16): p. 165016.
20. Li, Y. and S. Matej. *Deep Image Reconstruction for Reducing Limited-Angle Artifacts in a Dual-Panel TOF PET*. in *2020 IEEE Nuclear Science Symposium and Medical Imaging Conference (NSS/MIC)*. 2020. IEEE.
21. Sajedi, S., et al., *Intraoperative radio-guided imaging system for surgical applications*. 2019, Soc Nuclear Med.
22. Shopa, R.Y., et al., *Optimisation of the event-based TOF filtered back-projection for online imaging in total-body J-PET.* Medical Image Analysis, 2021. **73**: p. 102199.
23. Vandenberghe, S., et al., *Fast reconstruction of 3D time-of-flight PET data by axial rebinning and transverse mashing.* Physics in Medicine & Biology, 2006. **51**(6): p. 1603.
24. Worstell, W., et al., *Measurement of the parametrized single-photon response function of a large area picosecond photodetector for time-of-flight PET applications.* IEEE Transactions on Radiation and Plasma Medical Sciences, 2021. **5**(5): p. 651-661.
25. Flood, K., et al. *SymPET, a Waveform Digitizing" System on Chip" for Ultra-high Resolution TOF PET: Design Concept and Preliminary Studies*. in *2022 IEEE Nuclear Science Symposium and Medical Imaging Conference (NSS/MIC)*. 2022. IEEE.
26. Bläckberg, L., et al., *A layered single-side readout depth of interaction time-of-flight-PET detector.* Physics in Medicine & Biology, 2021. **66**(4): p. 045025.
27. Hashemi, A., et al. *VPG4, Versatile Parallelizable Geant4 Interface: A Novel Platform for Modeling Complex Nuclear Medicine Imaging Scanners*. in *2023 IEEE Nuclear Science Symposium, Medical Imaging Conference and International Symposium on Room-Temperature Semiconductor Detectors (NSS MIC RTSD)*. 2023. IEEE.
28. Hashemi, A., Y. Feng, and H. Sabet, *Equivariant Spherical CNN for Data Efficient and High-Performance Medical Image Processing.* arXiv preprint arXiv:2307.03298, 2023.
29. Hashemi, A., Y. Feng, and H. Sabet. *SCNN-Assisted Fast Image Reconstruction using Small Dataset: Gains and Limitations*. in *2023 IEEE Nuclear Science Symposium, Medical Imaging Conference and International Symposium on Room-Temperature Semiconductor Detectors (NSS MIC RTSD)*. 2023. IEEE.